\documentclass[aps,10pt,prl,tightenlines,twocolumn,twoside,showpacs,superscriptaddress]{revtex4-1}
\usepackage{amsfonts}
\usepackage{float}
\usepackage{placeins}
\usepackage{graphicx}
\usepackage[hidelinks]{hyperref}
\usepackage{color,amsmath,amssymb,bm}
\usepackage{tensor}

\usepackage[normalem]{ulem}  

\renewcommand{\sout}{\bgroup \color{red} \ULdepth=-.5ex \ULset}

\def\blfootnote{\xdef\@thefnmark{}\@footnotetext}

\newcommand{\beq}{\begin{equation}}
\newcommand{\eeq}{\end{equation}}
\newcommand{\bea}{\begin{eqnarray}}
\newcommand{\eea}{\end{eqnarray}}

\begin{document}

\title{
Spin Polarizations in a Covariant Angular-Momentum-Conserved \\Chiral Transport Model
}
\author{Shuai Y.F.~Liu\footnote{lshphy@gmail.com}}
\affiliation{Cyclotron Institute and Department of Physics and Astronomy, Texas A\&M University, College Station, Texas 77843-3366, USA}
\affiliation{Quark Matter Research Center, Institute of Modern Physics,
Chinese Academy of Sciences, Lanzhou 730000, China}
\author{Yifeng Sun\footnote{sunyfphy@gmail.com}}
\affiliation{Laboratori Nazionali del Sud, INFN-LNS, Via S. Sofia 62, I-95123 Catania, Italy} 
\author{Che Ming Ko\footnote{ko@comp.tamu.edu}}
\affiliation{Cyclotron Institute and Department of Physics and Astronomy, Texas A\&M University, College Station, Texas 77843-3366, USA}

\date{\today}

\begin{abstract}
Using a covariant and angular-momentum-conserved chiral transport model, which takes into account the spin-orbit interactions of chiral fermions in their scatterings via the side jumps, we study the quark spin polarization in quark matter.  For a system of rotating and unpolarized massless quarks in an expanding box, we find that side jumps can dynamically polarize the quark spin and result in a final quark spin polarization consistent with that of thermally equilibrated massless quarks in a self-consistent vorticity field.  For the quark matter produced in noncentral relativistic heavy ion collisions, we find that in the medium rest frame both the quark local spin polarizations in the direction perpendicular to the reaction plane and along the longitudinal beam direction show an azimuthal angle dependence in the transverse plane similar to those observed in experiments for the Lambda hyperon. 
\end{abstract}

\maketitle


\textit{Introduction.---} 
The spin-orbit interaction of chiral fermions can give rise to many interesting phenomena in physics, for 
example, the anomalous Hall effect~\cite{RevModPhys.82.1539} and the spin Hall 
effect~\cite{RevModPhys.87.1213,PhysRevLett.95.226801} in condensed matter systems as 
well as the Lambda hyperon spin polarization~\cite{Liang:2004ph,STAR:2017ckg}, the chiral vortical effect 
(CVE), and the chiral magnetic effect 
(CME)~\cite{Vilenkin:1979ui,Vilenkin:1980fu,Nielsen:1983rb,Fukushima:2008xe,Kharzeev:2015znc} in 
relativistic heavy ion collisions. Although the total angular momentum of a system of interacting chiral fermions 
is well defined, the separation of its spin and orbital contributions is subtle. As a result, various puzzles have 
arisen in understanding the spins of many physical systems, such as the origin of the proton 
spin~\cite{Ashman:1989ig,Ji:1996ek} as well as the relation between spin and orbital angular momentum in 
photon and laser physics~\cite{Leader:2013jra,Leader:2015vwa,Bliokh:2015doa} and twisted electron 
beams~\cite{Bliokh:2007ec,Barnett:2017wrr,Silenko:2017fvf}. Recently, the dependence of local lambda spin polarization on the azimuthal angle in the transverse plane of a heavy ion collision has been measured by the STAR Collaboration at the Relativistic Heavy Ion Collider (RHIC)~\cite{Adam:2019srw}, and it is found to disagree with theoretical predictions~\cite{Becattini:2017gcx,Xia:2018tes} based on the effect from the local thermal vorticity~\cite{Becattini:2013fla,Pang:2016igs}. In particular, the experimental and theoretical results have opposite sign in the azimuthal angle dependence, leading thus to the spin puzzle in relativistic heavy-ion collisions. This has since attracted much attention~\cite{Becattini:2017gcx,Huang:2019uhf,Florkowski:2019qdp,Wu:2019eyi,Becattini:2018lge,PhysRevC.100.064904,PhysRevC.101.024907,Sun:2018bjl}, including the works in Refs.~\cite{Becattini:2019ntv,Xia:2019fjf} 
that the sign problem cannot be explained by the effect from feeddowns of other hyperon states.

There are a number of general questions in the study of the spin dynamics of interacting chiral fermions. These include finding the right way to define their spins, the construction of a covariant description that respects the total angular momentum conservation, and the understanding of the anomalous effects induced by the spin-orbit interaction, such as the CVE, on the spin dynamics.  Studies along these directions have already been started by developing an angular-momentum-conserved relativistic hydrodynamics~\cite{Florkowski:2017ruc,Hattori:2019lfp}. In the present study, we make use of the widely studied chiral kinetic theory~\cite{Stephanov:2012ki,Son:2012wh,Gao:2012ix,Manuel:2014dza,Sun:2016nig,Huang:2018wdl,Zhou:2019jag}, especially the covariant chiral fermion theory developed in Refs.~\cite{Chen:2014cla,Chen:2015gta,PhysRevD.95.091901,PhysRevD.97.016004} for treating the scattering of two chiral fermions, to construct an angular-momentum-conserved chiral kinetic transport model and use it to address the spin puzzle in relativistic heavy ion collisions. 

\textit{The side jump formalism.---} 
As known in condensed matter systems~\cite{PhysRevB.2.4559}, to conserve the total angular momentum in a scattering process requires the inclusion of anomalous side jumps. The covariant formulation of this effect was developed in Refs.~\cite{Chen:2014cla,Chen:2015gta,PhysRevD.95.091901,PhysRevD.97.016004} for relativistic chiral fermions and shown to reproduce the chiral vortical effect in such systems~\cite{Vilenkin:1979ui}. These studies are, however, restricted to the scattering of particles with the same helicity at zero impact parameter and thus are not suitable for studying the spin-orbit interaction in relativistic heavy ion collisions. To go beyond this limitation, we follow Ref.~\cite{Chen:2015gta} by introducing a covariant four-dimensional angular momentum tensor for a particle, that is
\begin{align}
J^{\mu\nu}=x^{\mu}p^{\nu}-x^{\nu}p^{\mu}+S^{\mu\nu},
\label{eq_j1}
\end{align}
where $x$ and $p$ are, respectively, its four coordinate and four momentum, and $S^{uv} $ is its spin tensor. For a massless spin 1/2 particle, $S^{uv} $ can be chosen to be~\cite{Chen:2015gta}
\begin{align}
S^{\mu\nu}=\lambda{\frac{\epsilon^{\mu\nu\alpha\beta} p_\alpha n_\beta}{p\cdot n}} \,.
\label{eq_s1}
\end{align}
In the above, $n$ is the four vector that specifies the frame of reference; $ \lambda=\pm 1/2$ denotes the helicity of the particle with the plus and minus signs for positive and negative helicities, respectively.  This choice ensures that the spin tensor satisfies the conditions $p_\mu S^{\mu\nu}=0$ for Lorentz covariance and $n_\nu S^{\mu\nu}=0$ due to gauge fixing. To have a definite helicity for the massless particle further requires $n^{\mu}={(1,\bf{0})}$ and thus $S_n^{ij}=\lambda\epsilon^{ijk}p^k/|\bf{p}|$.  Because of parallel spin and momentum directions due to its definite helicity, the orbital motion of the massless particle also determines the motion of its spin, thus resulting in a strong spin-orbit coupling.  
 
As shown in Ref.~\cite{Chen:2015gta}, to maintain the helicity of a massless particle in any frame $n^\prime$ requires the imposition of $n^\prime=(1,{\bf 0})$ and also the introduction of an anomalous Lorentz transformation under which $J'=\Lambda J\Lambda^{\text{T}} $ and $p'= \Lambda p$ but 
\begin{align}
	x'^{\mu}=\Lambda\indices{^\mu_\alpha} x^\alpha+\Delta^{\mu}_{{\tilde n}n'},
	\label{eq_jumptrans}
\end{align}
where the term $\Delta^{\mu}_{{\tilde n}n'} $ describes the side jump and is given by
\begin{align}
\Delta^{\mu}_{{\tilde n}n'}=\lambda \frac{\epsilon^{\mu\alpha\beta\gamma}p'_{\alpha}{\tilde n}_{\beta}n_{\gamma}'}{(p'\cdot {\tilde n})(p'\cdot n')}
\label{eq_jumpdx}
\end{align}
with ${\tilde n}^\mu=\Lambda\indices{^\mu_\alpha}(1,\textbf{0})_\alpha =\Lambda\indices{^\mu_0}$. 

The above anomalous Lorentz transformation makes it possible to treat in a covariant way the conservation of total angular momentum in a two-body scattering  even for collisions of particles of different helicities and at a finite impact parameter.  With the initial momenta $\textbf{p}_1= -\textbf{p}_2=\textbf{p}$ and final momenta $\textbf{p}^\prime_{1}= -\textbf{p}^\prime_{2}=\textbf{p}'$ of the two scattering particles in their center-of-mass (CM) frame, conservation of the spatial part of the total angular momentum tensor, $J_1+J_2=J_1^\prime+J_2^\prime$, is reduced to 
\begin{align}
\textbf{J}=\mathbf{\Delta}\times \textbf{p}+(\lambda_1-\lambda_2)\hat{\textbf{p}}=\mathbf{\Delta}'\times \textbf{p}^\prime+(\lambda_1-\lambda_2)\hat{\textbf{p}}^\prime
\label{eq_cmj}
\end{align}
with $\mathbf{\Delta}$=$ \textbf{x}_1-\textbf{x}_2 $ and $\mathbf{\Delta}^\prime$=$ \textbf{x}_1^\prime-\textbf{x}_2^\prime $.  Since $ \hat{\bf p}'\cdot {\bf J}=\hat{\bf p}\cdot {\bf J}=\lambda_1-\lambda_2 $  is a constant and $|\textbf{p}|=|\textbf{p}^\prime|$, the momentum ${\bf p}'$ can be obtained from ${\bf p}$ by a rotation of angle $\phi$ around the direction of the total angular momentum $ \textbf{J} $, that is 
\begin{align}
\textbf{p}'&=R_{\hat{\textbf{J}}}(\phi)\textbf{p}
\nonumber\\
&\equiv(\textbf{p}\cdot \hat{\textbf{J}})\hat{\textbf{J}}+[\textbf{p}-(\textbf{p}\cdot\hat{\textbf{J}})\hat{\textbf{J}}]\cos\phi+(\hat{\textbf{J}}\times \textbf{p})\sin\phi.
\label{eq_roatep}
\end{align}
The value of the rotation angle $\phi$ depends on the angular distribution of the scattering cross section. Similarly, the relative distance ${\bf \Delta}^\prime$ can be obtained from ${\bf \Delta}$ via
\begin{align}
\mathbf{\Delta}^\prime=R_{\hat{\textbf{J}}}(\phi)\mathbf{\Delta}+\xi\textbf{p}^\prime,
\label{eq_rotatedx}
\end{align}
where the second term represents the ``gauge" freedom to shift $ {\bf \Delta}^\prime$ along the momentum direction without changing the cross product in Eq.(\ref{eq_cmj}). For simplicity, we choose $ \xi=0 $ in the present study. Since $ \mathbf{\Delta}\neq\mathbf{\Delta}^\prime$ in general, there is a change or jump in the relative distance between the two particles before and after scattering in their CM frame, which is  absent for scattering at zero impact parameter~\cite{Chen:2014cla,Chen:2015gta}. 

For the time component of the total angular momentum tensor, its conservation leads to the relation 
\begin{align}\label{dis}
\textbf{X}^\prime=\textbf{X}-(\textbf{p}-\textbf{p}^\prime)dt/\sqrt{s}\,
\end{align}
between the CM coordinates of the two colliding particles $\textbf{X} =(\textbf{x}_1+\textbf{x}_2)/2$ and $\textbf{X}'=(\textbf{x}'_1+\textbf{x}'_2)/2$ in their CM frame before and after the scattering. In Eq.(\ref{dis}),  $dt =x^0_1-x^0_2$ is nonzero in the CM frame since the scattering occurs at the same time in the laboratory (LAB) frame for the two particles, and $\sqrt{s}$ is the invariant energy of the two colliding particles in their CM frame.  With the scattering in the CM frame being local in time, one has $ x^{\prime 0}_1=x^0_1$ and $ x^{\prime 0}_2=x^0_2 $ and thus $dt'=x^{\prime 0}_1-x^{\prime 0}_2=dt$.  The coordinates of the two particles after scattering in the CM frame are then given by $ \textbf{x}^\prime_1=\textbf{X}^\prime+\mathbf{\Delta}^\prime/2 $ and $ \textbf{x}^\prime_2=\textbf{X}^\prime-\mathbf{\Delta}^\prime /2$.  Using Eqs.~(\ref{eq_jumptrans}) and (\ref{eq_jumpdx}), the coordinates and momenta of the two particles in the LAB frame can be obtained by a Lorentz boost with ${\tilde n}=(p_1+p_2)/\sqrt{s}$.  Since the time of each particle in the LAB frame changes after the scattering, there is an additional shift in its coordinate due to free propagation during this time change. The above procedure is time reversal invariant and guarantees the conservation of energy and momentum as well as the covariant total angular momentum in the LAB frame.

As shown in Ref.~\cite{Chen:2015gta}, side jumps lead to the conserved covariant current $j^{\mu}\approx p^\mu f+S^{\mu\nu}\partial_\nu f$, where $f$ is the particle phase space distribution, if one neglects the jump current due to scatterings of massless particles, which is small in a quark matter that is close to equilibrium as considered in the present study. In this case, the spin polarization densities can be expressed in terms of the phase-space distributions $f_{R/L}$ of left- and right-handed particles as
\begin{eqnarray}
j^{\mu}_{R/L}(x)=\int\frac{d^3p}{(2\pi)^3p}(p^\mu f_{R/L}+S^{\mu\nu}\partial_{\nu}f_{R/L} ).
\label{eq_jx}
\end{eqnarray}
In the above, the first term is the usual contribution $ \textbf{S}_\lambda=(2\pi)^{-3} \int d^3xd^3p\lambda \hat{\textbf{p}} f_\lambda(p,x)\equiv  \sum_i \lambda \hat{\textbf{p}}_i$ from the spin of a particle, while the second term is the magnetization contribution at $\hbar$ order~\cite{Chen:2014cla,Chen:2015gta} required by the covariance of the current density. Since the latter contribution is proportional to the vorticity field generated from the orbital motions of particles, it can be considered as the contribution from the local orbital angular momentum. We note that the spin tensor used in the present study and that in the widely used approach based on the Pauli-Lubanski pseudovector~\cite{Becattini:2013fla,Becattini:2017gcx} give the same spin polarization when global equilibrium is achieved but different for massless particles in local equilibrium.  According to Ref.~\cite{Becattini:2018duy}, finding the right spin tensor for particles in local equilibrium requires the comparison of its predictions with experimental observations.

Using the vector and axial currents given respectively by $j^{\mu}=j^\mu_{R}+j^\mu_{L}=(n, \textbf{j}) $ and $ j^{\mu}_5=j^\mu_{R}-j^\mu_{L}=(n_5, \textbf{j}_5)$, the distribution of spin polarization $ 2(\textbf{S}_R+\textbf{S}_L)/(N_R+N_L) $ in the system, where $N_R$ and $N_L$ are, respectively, the numbers of right- and left-handed particles, can then be expressed as
\begin{align}\label{pol}
\mathcal{P}=\int d^3x\textbf{j}_5(x)/\int  d^3x \,n(x)\,.
\end{align}
It is worthy to point out that defining the spin polarization through the covariant currents $j^\mu$ and $ j_5^\mu $ makes it possible to study its values in different frames of reference.  

\textit{Numerical implementations and results.---} The above theoretical framework can be implemented in the transport model by extending the usual parton scattering to include the side jumps to conserve the total angular momentum. We choose an isotropic cross section of 10 mb to mimic a strongly coupled medium and use the geometric method to determine the conditions for a scattering~\cite{Bertsch:1988ik,Li:2016uvu}.  To achieve the correct equilibrium distribution, the coordinates $ x_{1,2} $ of the two particles in the conditions for their scattering in the LAB frame are, however, modified by the side jump to $x_{1,2}-\Delta_{1,2}$.  Using the anomalous Lorentz boost of Eq.~(\ref{eq_jumptrans}), the scattering can also be treated in the CM frame using the geometric method with appropriate conditions. 

To illustrate the effect of side jumps on the scattering of massless quarks, we consider the case of a quark matter initially uniformly distributed in a box of sizes $ 5\times 5\times 5~\text{fm}^3$. For their initial momenta, they are taken to have a Boltzmann distribution at a temperature of $300$ MeV and a flow profile $ \gamma {\bf v}=2{\boldsymbol\omega} \times (x,0,0)$, where $ {\bf v} $ is the flow velocity, $\gamma=1/\sqrt{1-v^2}$ is the corresponding gamma factor, and $\boldsymbol\omega=(0,0.12,0)$ $({\rm fm}/c)^{-1}$ is a constant vorticity field along the $y$ direction.  The system is then allowed to expand for 2 fm/$c$.

\begin{figure} [!htb]
	\centering
	\includegraphics[width=0.99\columnwidth]{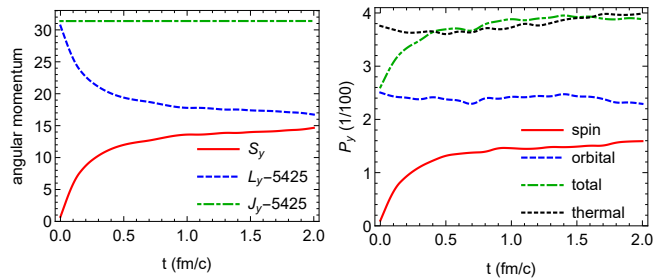}
	\caption{Left panel: Quark spin $S_y$, orbital $L_y$, and total $J_y$ angular momenta in the $y$ direction. Right panel: Quark total spin polarization in the $y$ direction and its spin and orbital contributions together with the spin polarization expected from the thermal model.}
	\label{fig_bench}
\end{figure}

In the left window of Fig.~\ref{fig_bench}, we show the time evolution of the spin $S_y$ (solid), orbital $L_y$ (dashed line) and total $J_y$ (dash-dotted line) angular momenta of the system along the $y$ direction with the orbital and total angular momenta subtracted by the constant value of 5,425$\hbar$. It is seen that the spin angular momentum increases from zero and quickly reaches 12$\hbar$ within 0.5 fm/$c$.  The orbital angular momentum decreases by the same amount, and the total angular momentum of the system is conserved during the evolution. The total polarization in the $y$ direction (dash-dotted line) given by the sum of the spin (solid line) and orbital (dashed line) terms in Eq.~(\ref{eq_jx}) is shown in the right window of Fig.~\ref{fig_bench} and compared to the thermally equilibrated value $\omega/(2T)$ (dotted line), where both $ \omega $ and $ T $ are calculated by assuming quarks are in local thermal equilibrium. In calculating the average polarization, we have only used the volume $ 3\times 3\times 3$ fm$^3$ around the center of the system to avoid numerical complications near the boundary of the expanding box.  We see that the contribution to the total polarization from the spin term in Eq.~(\ref{eq_jx}) increases from zero and quickly reaches a plateau as in the behavior of the total spin.  For the contribution from the orbital term in Eq.~(\ref{eq_jx}), it is finite even at $t=0$ due to local orbital motions as a result of the nonzero vorticity field. As to the total spin polarization, its final value is close to that expected from the thermal model result.

We further study the quark spin polarizations in heavy ion collisions by considering Au + Au collisions at $ \sqrt{s}=200$~GeV and at impact parameter $ b= 8.87$~fm, corresponding to the collision centrality of 30\%-40\%.  For the initial quark and antiquark phase-space distributions, we take them from a multiphase transport (AMPT) model with string meltings~\cite{Lin:2004en}.  The resulting partonic matter is then evolved according to the above described covariant and angular-momentum-conserved transport model. Because the quark polarization is related to the axial current $j_5^{\mu}$, which is a Lorentz four-vector, the values of its components depend on the frame of reference where they are evaluated. With the flow velocity given by $ \textbf{v}(x)=\textbf{j}(x)/n(x)$ as well as the relations $ \int  d^3x' \,n'(x)=\int  d^3x \,n(x) $ and $ \int  d^3x' \,j_5'(x)=\int  \gamma d^3x \,\Lambda j_5(x) $ between quantities in the local medium rest frame and the LAB frame, the quark spin polarizations in the medium rest frame can then be evaluated  from the spin polarization in the LAB frame using Eq.(\ref{pol}).   As in the experimental measurement of Lambda hyperon polarization~\cite{STAR:2017ckg,Adam:2019srw}, we use the event plane determined from the elliptic flow of partons to define the azimuthal angle $\phi$ in the transverse plane of a heavy ion collision. To mimic the pseudorapidity cut $-1<\eta<1$ in experiments, we only integrate the spatial region  where the flow velocity is in the pseudorapidity range $-1<\frac{1}{2}\rm{ln}\frac{v+v_z}{v-v_z}<1$.

\begin{figure}[h]
	\centering
	\includegraphics[width=1.00\columnwidth]{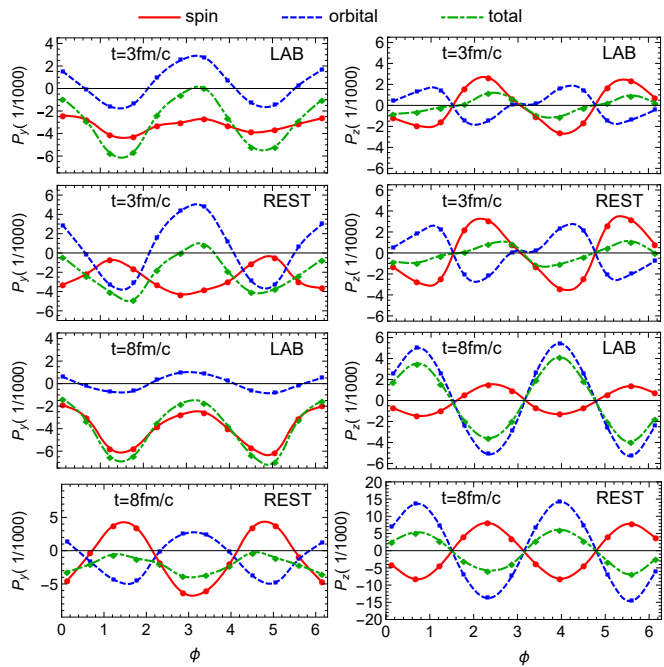}
	\vspace{-0.7cm}
	\caption{Azimuthal angle $\phi$ dependence of spin polarizations along the $y$ direction and $z$ direction in the LAB and medium rest (REST) frame at times $t=3$ and 8 fm$/c$.
	}
	\label{fig_pypz}
\end{figure}

Figure~\ref{fig_pypz} shows the azimuthal angle distribution in the transverse plane for local quark spin polarizations $\mathcal{P}_y$ along the $y$ direction perpendicular to the event plane (left windows) and $\mathcal{P}_z$ along the longitudinal $z$ direction (right windows) in the LAB frame  and the medium rest frame.  For the total spin polarization $\mathcal{P}_y$ (dash-dotted line) along the $y$ direction, both the spin (solid line) and orbital (dashed line) contributions have the similar azimuthal angle dependence $-a+b\cos 2\phi$ ($a,b>0$) in the LAB frame with the former having a larger magnitude.  After transforming to the medium rest frame, the azimuthal angle dependence of the orbital contribution remains the same except having an enhanced magnitude but the spin contribution changes its azimuthal angle dependence to $-a-b\cos 2\phi$ ($a,b>0$). For the spin polarization $\mathcal{P}_z $ along the $z$ direction, its spin and orbital contributions have the azimuthal angle dependence $\sin 2\phi$ and $-\sin 2\phi$, respectively, in the LAB frame. The magnitudes of both spin and orbital contributions are enhanced after transforming to the medium rest frame. As to the relative strength of the orbital vs spin contribution, it changes with time. For $P_y$, it is dominated by the orbital contribution at earlier times and by the spin contribution at later times. This behavior is reversed for $P_z$ as it is dominated by the spin contribution at earlier times and by the orbital contribution at later times.  We note that our results at earlier times are close to those predicted by the thermal model~\cite{Becattini:2017gcx}. The azimuthal angle dependence of the quark spin polarizations along the $y$ and $z$ directions in the medium rest frame obtained in the above appears to be similar to that of the Lambda hyperon measured in its rest frame~\cite{Niida:2018hfw,Adam:2019srw}.  Since the spin of the Lambda hyperon in the constituent quark model is given by its strange quark, the azimuthal angle dependence of the local spin polarizations of Lambda hyperons in their rest frames and strange quarks in the medium rest frame in relativistic heavy ion collisions are closely related.  Our model thus provides a promising framework for understanding the spin polarization of Lambda hyperon measured in RHIC experiments. 

\begin{figure} [!thb]
	\centering
	\includegraphics[width=1.0\columnwidth]{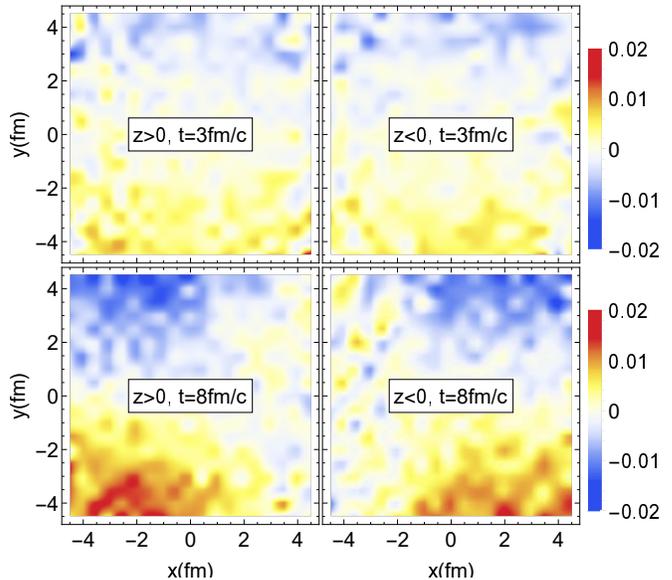}
	\caption{The distribution in the transverse plane of final LAB frame for the integrated axial charge to charge ratio at times $t=3$ and 8 fm/$c$ in the regions of $z>0$ (left panels) and $z<0$ (right panels)  along the longitudinal beam direction.}
	\label{fig_n5}
\end{figure}

These results can be understood from the evolution of the axial charge distribution in the quark matter. In the chiral kinetic theory, the axial charge is the time component of the four-dimensional axial current $ j^\mu_5 $ and plays an important role in the relativistic spin transport.  It was shown in Ref.~\cite{Sun:2018bjl} that the redistribution of axial charges in the transverse plane of heavy ion collisions is essential for determining the azimuthal angle dependence of the quark spin polarizations.  In Fig.~\ref{fig_n5}, we show the results obtained in the present study for the distribution in the transverse plane of the ratio of the final integrated axial charge density $\int dz\, n_5(x,y,z)$ to the total particle density $\int dz\, n(x,y,z)$ calculated in the LAB frame for the regions of $z>0$ (left panels) and $z<0$ (right panels).  At the early time of 3 fm$/c$, small net axial charges of opposite signs are seen to appear in the positive and negative $y$ regions. The above axial charge dipole moment in the transverse plane changes to a quadrupole distribution at the later time of 8 fm$/c$.  Since the axial charge current transformed under the Lorentz boost of velocity $\textbf{v}$ according to $\textbf{j}'_5=\textbf{j}_5+(\gamma -1)(\textbf{j}_5\cdot\hat{v})\hat{v}-\gamma n_5\textbf{v}$, the quark spin polarization can have different values at different frames of reference due to the redistribution of $ n_5 $. For example, the flow velocity $ v $ at the azimuthal angle $\phi=\pi/2$ has a positive $ y $ component where the axial charge $ n_5 $ is negative. Thus, the Lorentz boost can increase the spin part of $ \mathcal{P}_y$ in the medium rest frame and changes its azimuthal angle dependence. 

Above results are obtained with a large quark scattering cross section required by the large elliptic flow observed in experiments as shown in Refs.~\cite{Sun:2016nig,Sun:2016mvh}.  With a smaller cross section, the total quark spin polarization would decrease and the azimuthal angle dependence of its longitudinal polarization would also change because of the weakened redistribution of axial charges in the transverse plane.

\textit{Conclusions.---} 
Based on the side-jump formalism for the scattering of chiral fermions, we have constructed a transport model that conserves the total angular momentum of the quark matter created in relativistic heavy-ion collisions.  Via the introduction of a covariant angular momentum tensor, the spin polarization of massless quarks then has both spin and orbital contributions. In the case of an expanding box with a given vorticity field, we have found that the final quark spin polarization is consistent with that expected from the thermal model. For the case of a heavy ion collision using initial conditions from the realistic AMPT model, we have further found that the quark local spin polarizations can be understood from the redistribution of the axial charges induced by CVE, leading to the appearance of dipole and quadrupole structures in the transverse plane of heavy ion collisions. Including both the spin and orbital contributions to the quark local spin polarizations, we have found that their azimuthal angle dependences along the transverse and longitudinal directions in the medium rest frame are similar to those of Lambda hyperons observed in experiments. 

For a quantitative understanding of the sign problem in the $\Lambda$ hyperon local spin polarization requires further studies.  First, for strange quarks, an improved approach needs to be developed to include their finite masses in their angular-momentum-conserved scatterings, where the theoretical framework has been studied in Ref.~\cite{Li:2020vwh}. Also, this massive scattering process should be incorporated in a quasiparticle transport model to properly take into account the equation of state of quark matter. Furthermore, it is important to understand how to form a $ \Lambda $ hyperon from its constituent quarks in an angular-momentum-conserved coalescence approach that includes the anomalous contribution to its spin and polarization. Future studies along these directions will also provide the theoretical framework to understand more generally the anomalous transport phenomena in quark matter from experiments on heavy ion collisions. 

\acknowledgments

This work was supported in part by the US Department of Energy under Contract No. DE-SC0015266 and the Welch Foundation under Grant No. A-1358.
\bibliography{refcnew}
\end{document}